# The electric field of the Earth after the occurrence of the February 14th, 2008, Ms = 6.7R EQ in Greece. Its implications towards the prediction of a probable future large EQ.

C.P.Thanassoulas[1] B.Sc in Physics, M.Sc – Ph.D in Applied Geophysics

1. Retired from the Institute for Geology and Mineral Exploration (IGME)
Geophysical Department, Athens, Greece.
e-mail: thandin@otenet.gr

The electric field of the Earth registered by three monitoring sites (ATH, PYR, HIO) located in Greece, is investigated and analyzed after the occurrence of the Methoni EQ (14th of February, 2008, Ms =6.7R). The period of analysis is performed for 2 days (21st-22nd of February) and 7 days after the occurrence of the main seismic event. The obtained results suggest that the seismogenic area generates electrical signals denoting a specific epicentral area. This area coincides with the already seismically activated area. An estimate for the time of occurrence of this EQ is made by the application of the Oscillating Lithospheric Plate Model. The analysis of the seismic potential of the regional area suggests that the remaining stored seismic energy is capable of producing a large earthquake in the same area. The expected maximum magnitude (Ms) of a future earthquake which could take place in the same seismogenic area is estimated as Ms = 7.24R by the application of the Lithospheric Seismic Energy flow model.

## 1. INTRODUCTION

On 14th of February, 2008, a large (Ms =6.7R) EQ took place in a short distance (about 30Km) from Methoni town of Southern Peloponnese. This EQ was the third in the row in a short time period out of a total of three which occurred in the regional area of Peloponnese, Southern Greece. All three EQs are presented in the following figure (1). The first EQ (Leonidion, January 6th, 2008, Ms = 6.6R) is denoted by the letter A, the second EQ (Methana, January 29th, 2008, Ms = 5.1R) is denoted by the letter B and the third one (Methoni EQ, February 14th 2008, Ms = 6.7R) is denoted by the letter C.

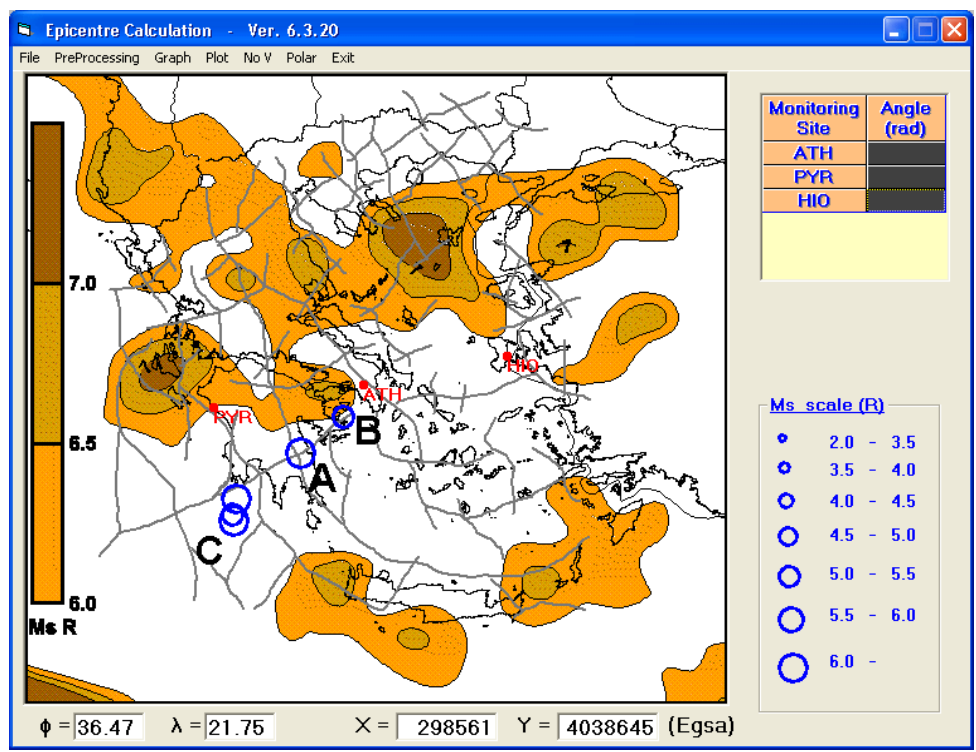

Fig. 1. Location of the epicentral areas of Leonidion (A), Methana (B) and Methoni (C) EQs.

It must be pointed out that all three (A, B, C) epicenters are located along the corresponding lithospheric fracture zone (gray thick line) which strikes NE – SW at the same regional area. In the epicentral area C two more circles are shown which represent largest aftershocks (Ms>6R) in the same seismogenic area. In the same map of figure (1) the PYR, ATH and HIO monitoring sites location is represented by a red dot and red lettering. These three monitoring sites form a linear array of electrical field receivers which presents a wide angle azimuth reception (main receiving wide lobes directed at N – S direction) thus detecting electrical signals which could be generated at any part of Greece. It is understood that minimal array reception occurs along the array direction (almost E – W). More details about the specific EQs of Leonidion and Methoni can be found in [2].

## 2. THE THEORETICAL MODELS

Generally, the problem of the earthquake prediction is considered by the seismologists as an unsolvable one. Their positions which defend the unpredictability of the EQs have been presented in various published papers. The main objection is that there is no physical model to support analytical solution of the three parametric (time, location and magnitude) prediction problem while on the other hand the generation of an earthquake is a random event which obeys more or less stochastic laws. Moreover, the so-called "short-term earthquake prediction" in terms of some days or hours in advance of the main large seismic event is completely considered as impossible.

However, in terms of Applied Geophysics, Rock Mechanics, Electric Potential Fields and Energy Conservation Law of Physics this unsolvable problem has a valid physical solution which has been verified from field experiments during the last 4 - 5 years (2003-2008). A detailed presentation of this methodology was presented by Thanassoulas [1] in a monograph under the title "Short-term Earthquake Prediction".

In the process of solving this problem three different physical models have been used.

<u>The first one is the oscillating lithospheric plate due to tidal forces</u>. Actually, this physical mechanism provides the last tiny extra amount of stress load which forces the seismogenic area to exceed its fracture stress level. Due to the nature of the tidal forces it is possible to determine the times when tidal stress load of the lithosphere reaches maxima. Therefore, generally, large earthquakes take place mostly during these tidal peaks. Statistical analysis of the time of occurrence of large EQs vs times of tidal peaks (with period of 1 day or 14 days) has shown [1] that it is possible to determine in favorable cases the time of occurrence of an EQ in a time window as short as less than an hour. This had been demonstrated for the Skyros EQ (26/7/2001, Ms=6.1R) in Greece, which had been announced 3 days in advance (23/7/2001) in a presentation of the specific methodology at the Bulgarian Academy of Sciences (BAS). A written verification of this prediction was produced by the BAS and is presented in the monograph as in [2] as well. This model, although provides a specific number of times when maxima of the lithospheric stress load take place it is not capable to identify the specific maximum when a large EQ will occur. Therefore, there is the necessity to narrow down the "time solutions" which are provided by the tidally triggered lithospheric oscillating plate by another set of solutions which will be derived by the application of a different physical model applied on an independent physical parameter of the lithosphere. Here comes the contribution of the second model based on Rock Mechanics and the Piezoelectricity.

<u>The second model takes the advantages of the theories of rock deformation under stress load and the corresponding generated piezoelectric phenomena associated with it.</u> The stress – strain curve that represents the deformation of any material under stress increase conditions is an elementary part of any Rock Mechanics text-book. In association to this mechanism is the generated piezoelectric potential which depends strictly on the piezoelectric properties of the material. In example, quartzite and tourmaline exhibit the largest piezoelectric properties from all rocks. Therefore, since the lithosphere presents a large amount of quartzite as its constituent, as a result by the combination of the oscillating lithospheric plate, in general, electrical oscillating signals must be generated due to the interaction of the piezoelectric rock formation of the lithosphere and its oscillation due to the effect of tidal waves. This phenomenon is intensified when the stress-load of the lithosphere reaches its maximum level so that maximum deformation exists just prior to fracturing. Consequently, an indicator that the last phase of deformation has been reached and that fracturing is imminent, is the fact of the presence of mainly oscillating electrical signals plus any other type (VLP, Spikes) generated by the pre-fracturing process. Examples of such signals which preceded large EQs were presented elsewhere [1]. As a result, the time solutions obtained from the application of the first model are narrowed down drastically. Therefore, the remaining solutions which obey both models lay in the area of days – hours or in other words are "short-term prediction" time solutions. Moreover, the presence of the electrical signals and specifically the ones of very long wavelength allow us to calculate the azimuth direction of the Earth's electric field intensity vector. This calculation allows us by triangulation to determine the location of the origin of the electrical field which is generated by the deformation of the seismogenic area.

So far, two parameters can be determined: the time of occurrence and the location of the epicentral area. What is left is the magnitude of the expected earthquake. This is dealt with a third different model.

<u>The third model assumes that the seismogenic area is considered as an open physical system.</u> This model is studied in terms of energy balance between the absorbed (due to plate motion and other mechanism) potential energy and the released one due to "normal" regional seismicity. This is achieved by converting all past earthquakes into released energy and by constructing the cumulative seismic energy release graph of the specific seismogenic region. The latter has been identified a priory by the triangulation procedure of the $2^{nd}$ model. The difference between the absorbed and the released energy is what is stored in the seismogenic region and may be released in a future EQ. The max magnitude of the EQ is calculated by inverting the "max to be released energy" in to the corresponding magnitude. In practice, the achieved accuracy [1] is of the order of dM = .07R compared to the dM = 0.5 achieved by seismological statistical methods.

All the latter will be applied on the data recorded on $21^{st}$ to $22^{nd}$ of February by the PYR, ATH and HIO monitoring sites in an attempt to analyze the status of the seismogenic region of Methoni EQ after 7 days from the occurrence of the main seismic event and for a period of two consecutive days.

## 3. THE DATA

At the monitoring sites (PYR, ATH, HIO) of the Earth's electric field continuous registration (at NS / EW directions) takes place at a rate of 30000 samples per second. The average value of each minute is stored in the hard disk of the data logger. For the last six months period the data are normalized to a unit dipole for each monitoring site. In the following figures 2, 3 and 4 the recordings which correspond to PYR, ATH and HIO monitoring sites are presented.

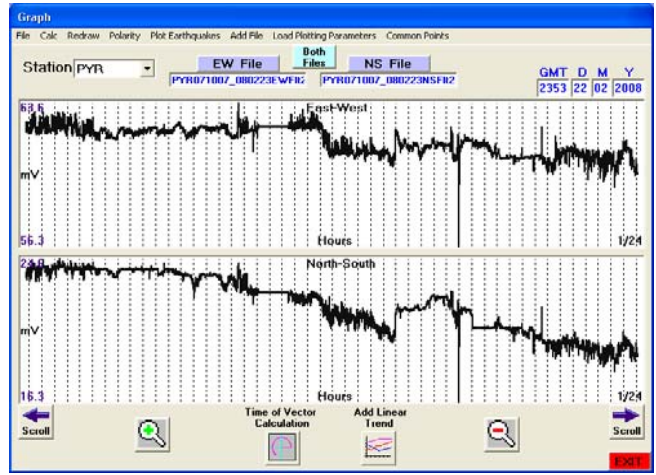

Fig. 2. Normalized values of the Earth's electric field recorded by PYR monitoring site for the period from 21st February to 22nd February 2008.

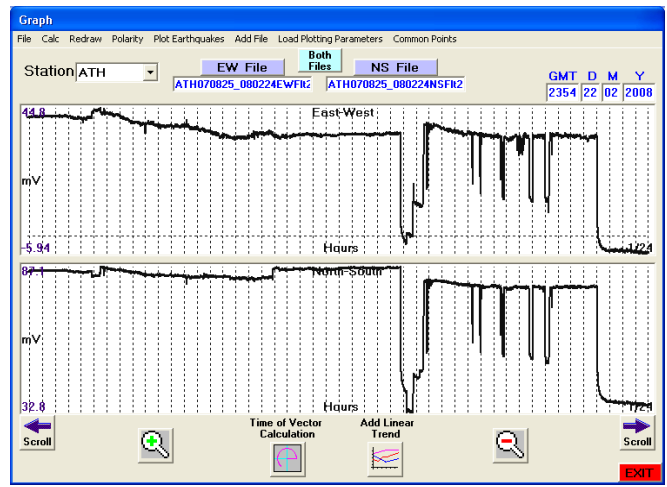

Fig. 3. Normalized values of the Earth's electric field recorded by ATH monitoring site for the period from 21st February to 22nd February 2008.

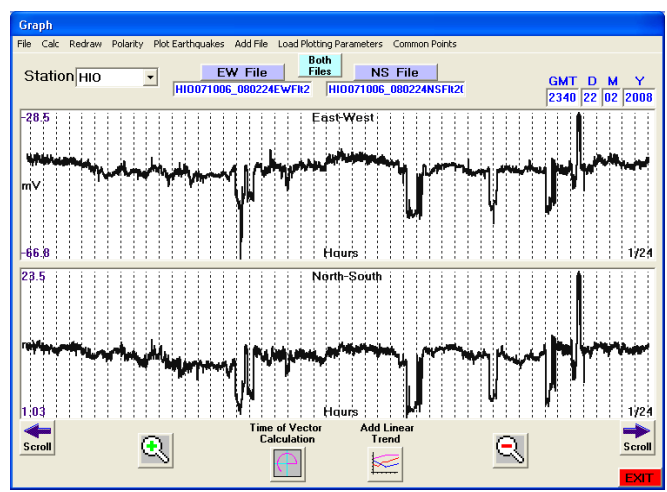

Fig. 4. Normalized values of the Earth's electric field recorded by HIO monitoring site for the period from 21st February to 22nd February 2008.

The vertical dashed lines indicate 1 our intervals, while the vertical scale is in mV.

## 4. DATA PROCESSING.

The analysis to follow will be performed in a monochromatic mode. The single period which was chosen to this end is the 24 hours daily oscillation. The main advantage of this specific period is that it coincides to the corresponding daily tidal oscillation of the lithospheric plate. If the model of the lithospheric plate is valid then this oscillating electric field will be strongly affected by the stress load of the future seismogenic region following the piezoelectric mechanism too. The detailed methodology to follow was presented by Thanassoulas [1].

The normalized data of figures 2, 3, and 4 were filtered in a band - pass mode by using the FFT methodology. The obtained results are presented in the following figures 5, 6, 7.

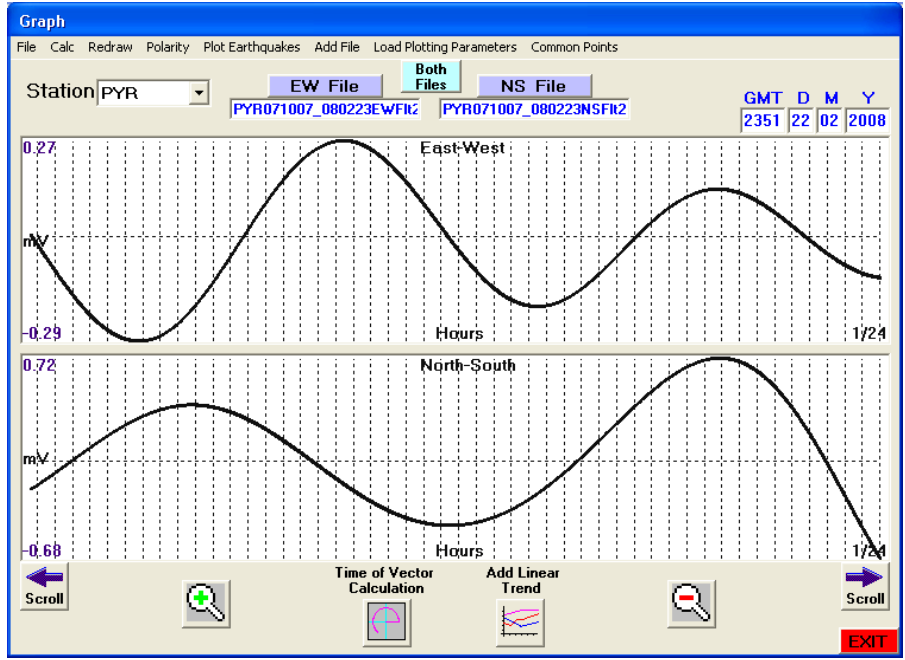

Fig. 5. Band - pass filtered data of PYR monitoring site. T = 1 day (24 hours).

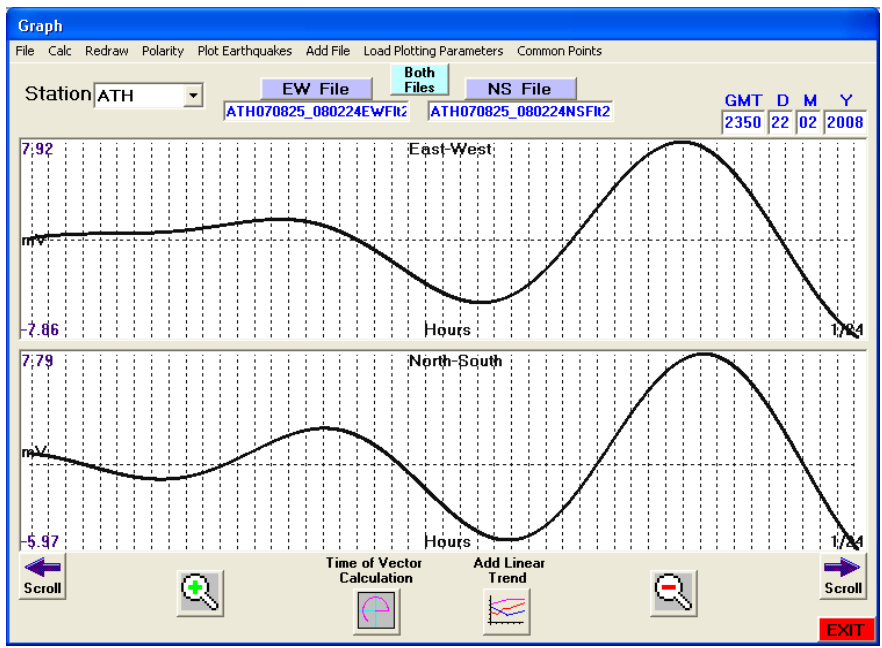

Fig. 6. Band - pass filtered data of ATH monitoring site. T = 1 day (24 hours).

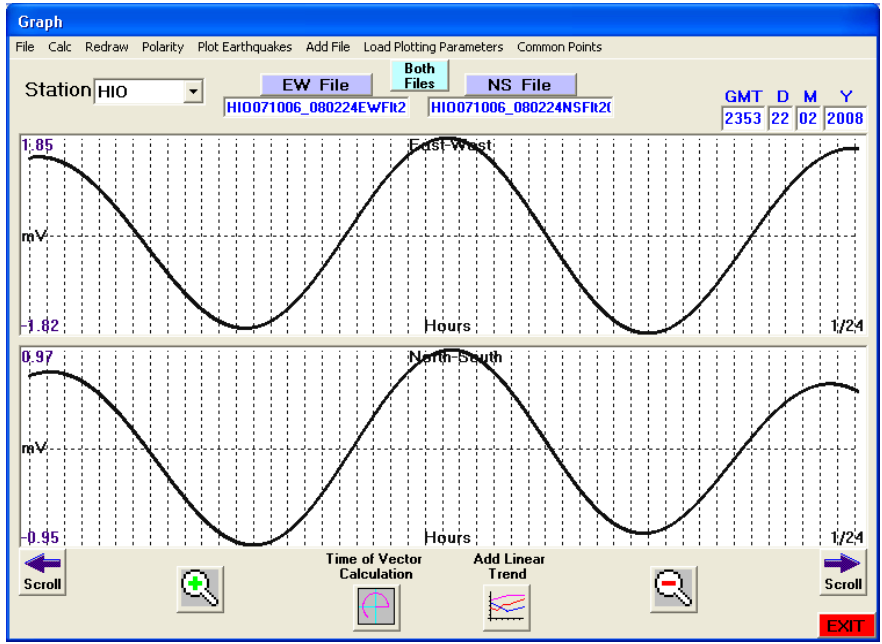

Fig. 7. Band - pass d filtered data of HIO monitoring site. T = 1 day (24 hours).

At this point for each monitoring site we have two orthogonal components (EW – NS) of an oscillating field. Therefore it is possible to determine for each sample point (1 minute interval) the azimuth direction of the electric field intensity vector. For each monitoring site its average azimuth value will indicate the azimuth direction of the origin of the source of the electric field in respect to the location of the same monitoring site. This procedure is a simple arctan trigonometric determination (in rad, using the normal trigonometric unit circle) for each pair (NS – EW) of sample values.

The obtained results (azimuth diagram) from each monitoring site along with its location and the corresponding map of Greece are presented in the following figures 8, 8a, 9, 9a, 10, 10a.

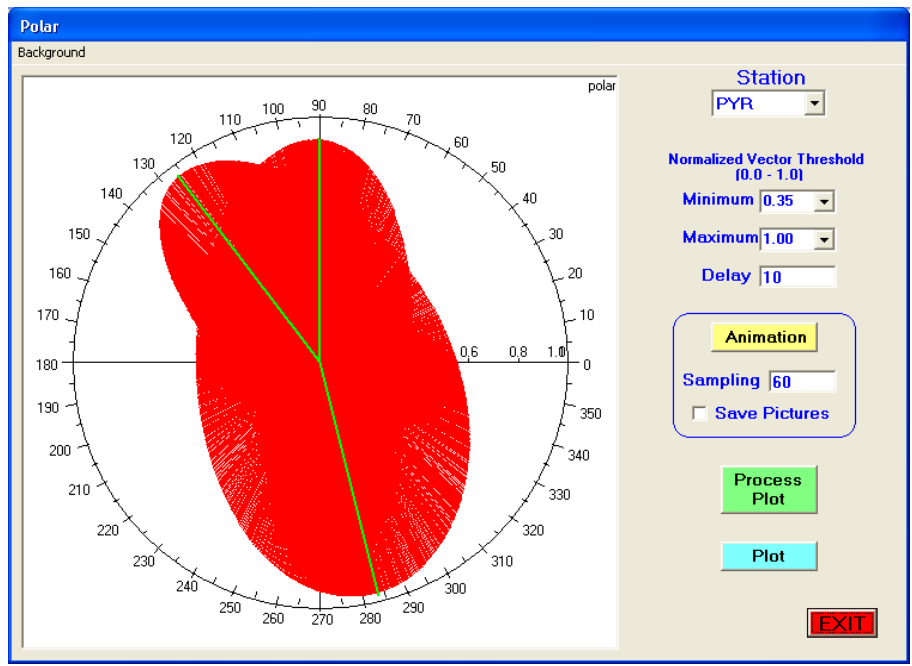

Fig. 8. Polar distribution of the Earth's electric field intensity vector determined for PYR monitoring site. The red colored lines (red area) represent each azimuth calculation for each sample point while the green line indicates the characteristic average azimuths of the entire two days signal analysis. Total number of samples = 2880. Determined azimuths = 4.95 / 4.80 rad.

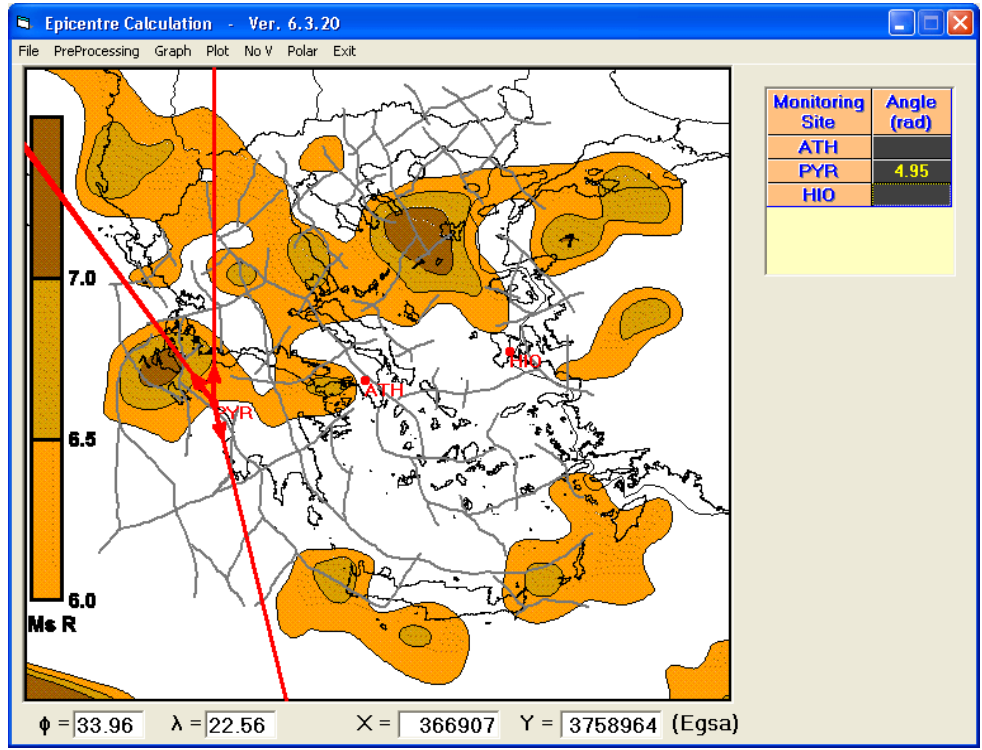

Fig. 8a. Azimuth directions (red lines) obtained from PYR monitoring site on top of map of Greece. Brown colors indicate the seismic potential distribution over Greece for the year 2000 [1].

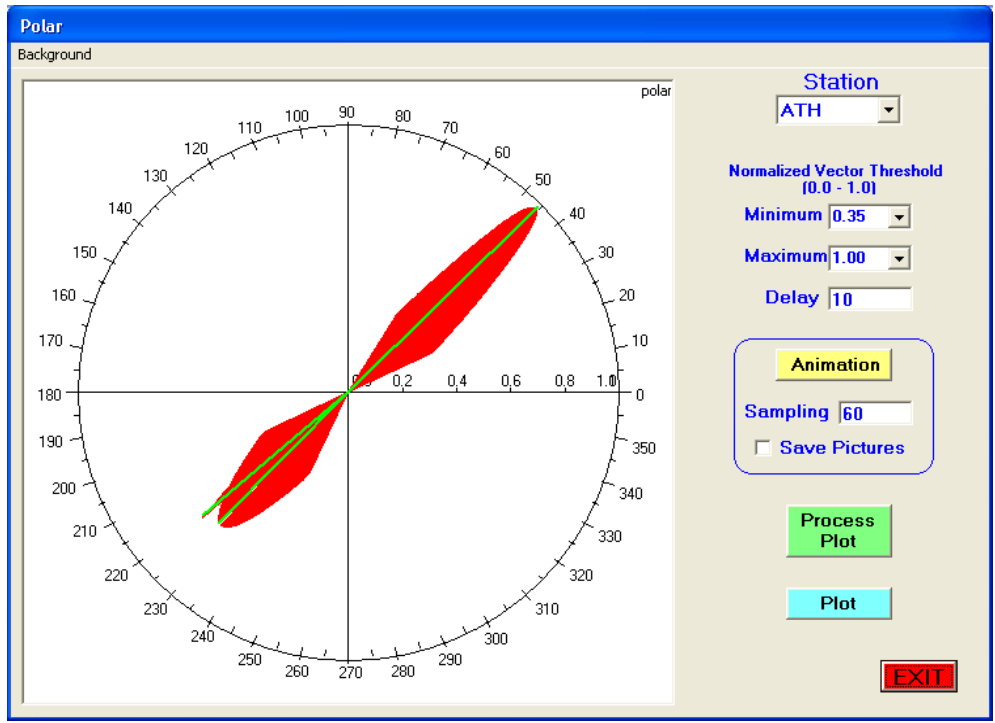

Fig. 9. Polar distribution of the Earth's electric field intensity vector determined for ATH monitoring site. The red colored lines (red area) represent each azimuth calculation for each sample point while the green line indicates the characteristic average azimuths of the entire two days signal analysis. Total number of samples = 2880. Determined azimuths = 3.86 / 3.94 rad.

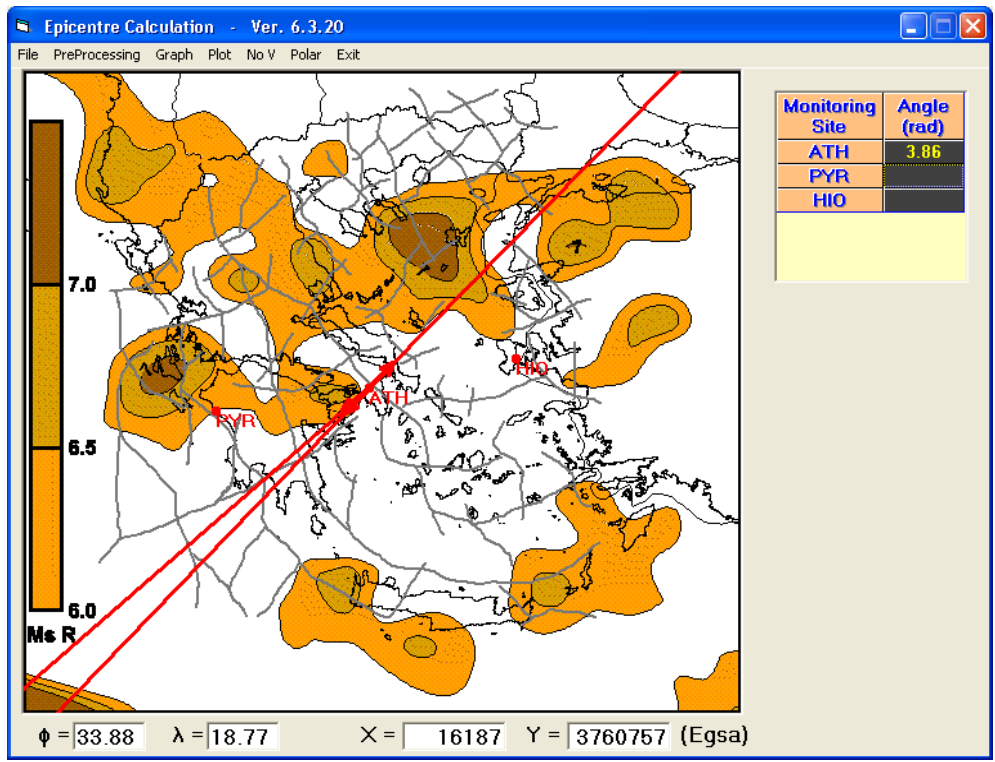

Fig. 9a. Azimuth directions (red lines) obtained from ATH monitoring site on top of map of Greece. Brown colors indicate the seismic potential distribution over Greece for the year 2000 [1].

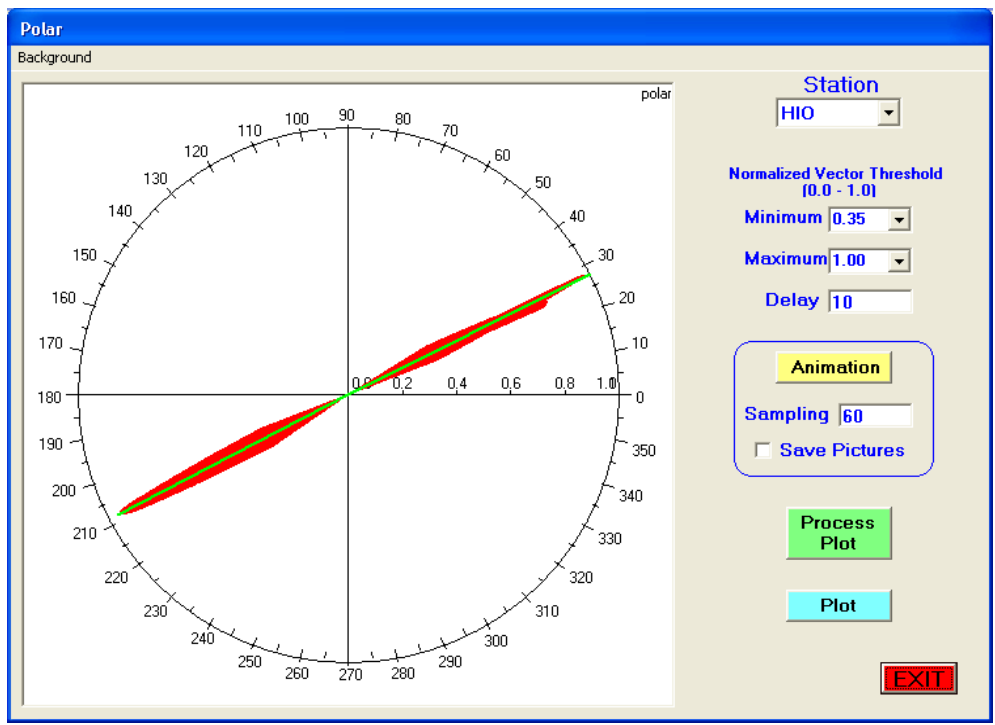

Fig. 10. Polar distribution of the Earth's electric field intensity vector determined for HIO monitoring site. The red colored lines (red area) represent each azimuth calculation for each sample point while the green line indicates the characteristic average azimuths of the entire two days signal analysis. Total number of samples = 2880. Determined azimuths = 3.64 rad.

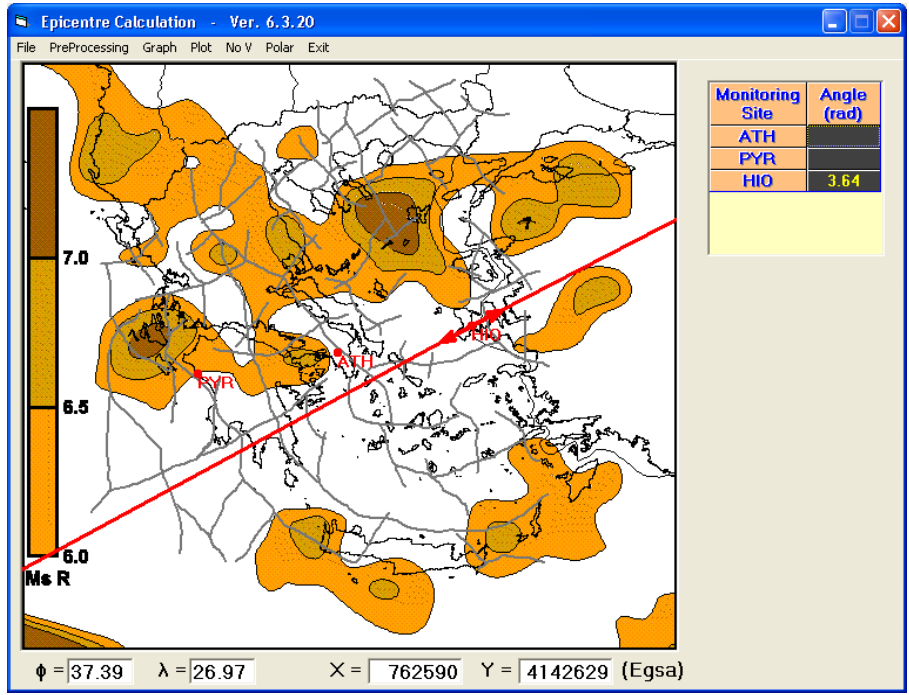

Fig. 10a. Azimuth directions (red lines) obtained from HIO monitoring site on top of map of Greece. Brown colors indicate the seismic potential distribution over Greece for the year 2000 [1].

For all cases the NE and NW azimuths were discarded since they do not converge. For the rest of them in cases where more than one azimuth directions were obtained but showing a small difference between each other the average value was considered as more representative. Therefore, the following values were adopted for each monitoring site:

PYR = (4.95+4.80)/2 = 4.87 rad,   ATH = (3.86+3.94)/2 = 3.90 rad,   HIO = 3.64

These average azimuth values were used to compile the following map which is presented in figure (11)

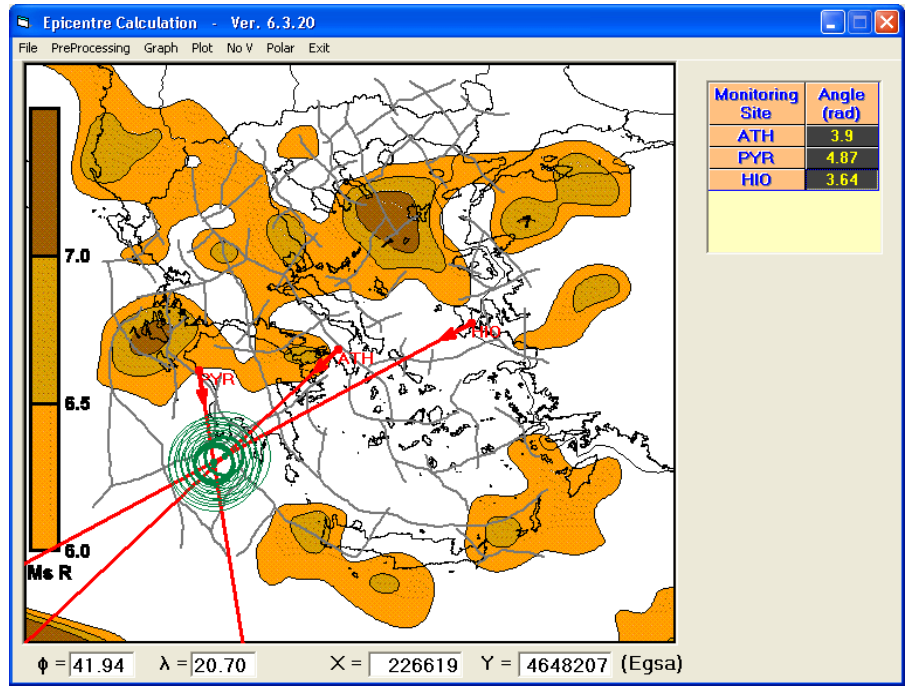

Fig. 11. Simultaneous presentation of all average azimuths obtained from each monitoring site (PYR, ATH, HIO). The green circles indicate the location of the origin of the electrical field generating mechanism. The intersection of the azimuth directions is made in two distinct geographical coordinates: a. 36.40N / 21.61E and b. 36.59N / 21.97E.

It is very interesting to compare the location of the Methoni seismic event at figure (1) with the calculated location of the origin (fig. 11) of the electrical field generating mechanism. The fit is more than satisfactory.

Up to this point it has been demonstrated that an electric field generated in the focal area of Methoni EQ is still present and detectable at large distances from Methoni EQ epicentral area. This fact suggests that the Methoni regional seismogenic area is still active and strain / stress charged. Therefore, since it is at critical strain charged conditions the question raised is as follows: if an earthquake is to occur in the future what is its most likely time of occurrence?

This question will be answered by the use of the oscillating lithospheric plate model. The answer is straightforward. Actually the future earthquake will occur most probably at the next maximum amplitude of the lithospheric oscillation. Firstly the 14 days period lithospheric oscillation is examined. This is presented in the following figure (12).

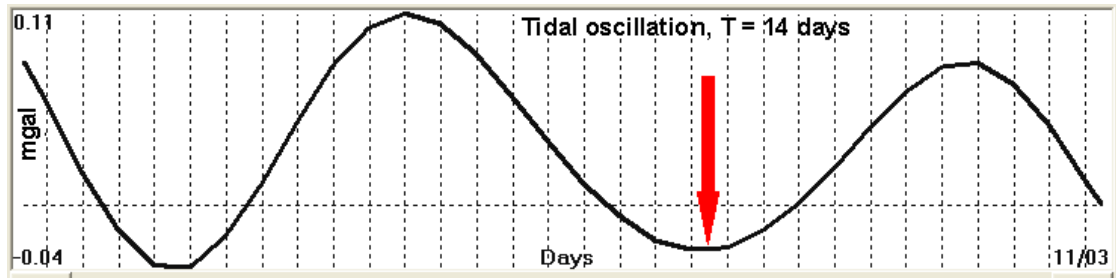

Fig. 12. 14 days period lithospheric oscillation (black line) determined for a period from 10/2/2008 to 11/3/2008. The local minimum corresponds to the 29$^{th}$ of February, 2008 (red arrow).

If an error of one day is accepted, then the probable future EQ must occur in the period from 28$^{th}$ 0f February to 1$^{st}$ of March. The time window for the occurrence of this EQ has been narrowed down to 3 days. Next it is possible to determine the most probable times within each of these three days by inspecting the daily tidal lithospheric oscillation. This is presented in the following figure (13).

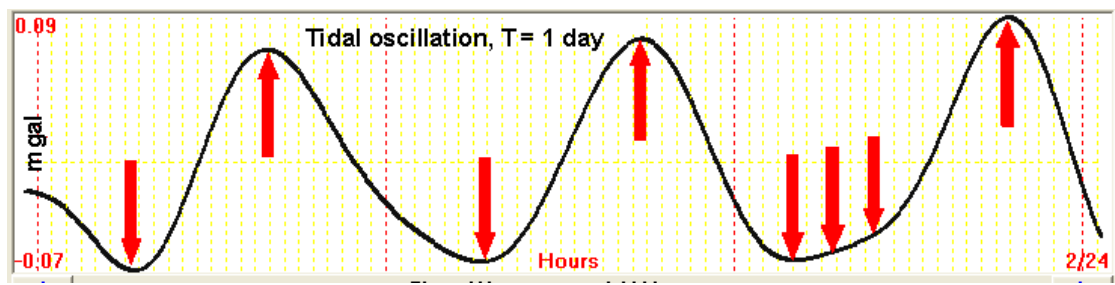

Figure 13. Daily lithospheric oscillation for the period from 28/2/2008 to 1/3/2008. The red arrows indicate local maxima and minima of the tidal lithospheric oscillation within each day.

Consequently, the suggested time of occurrence is as follows:

Day       : 28/2 – 1/3/2008

Time     :28/2/2008         : 06.30 or 15.50 GMT

            :29/2/2008         :06.30  or 17.36 GMT

            :01/3/2008         :04.17  to 09.44 or 18.57 GMT

For these specific times an error of +/- 1 hour is acceptable. The largest the EQ is the most accurate is this estimation. For the case of the East Kythira EQ (08/01/2006, Ms=6.9R) the time of occurrence coincided exactly with the day while there was an error of only 40 minutes from the corresponding daily tidal minimum.

Finally, in order to complete this work an estimate must be provided for the magnitude of the probable future EQ. This will be utilized by the use of the Lithospheric seismic energy flow model [1]. In this methodology, the regional seismogenic area is considered but taking into account the large lithospheric fracture zones which exist in the study area. This is demonstrated for this case in figure (14). The selected frame has been oriented towards a NE – SW direction following the corresponding lithospheric fracture zone. Moreover, the spatial distribution of the aftershocks of the Methoni EQ, show the same directional spatial distribution.

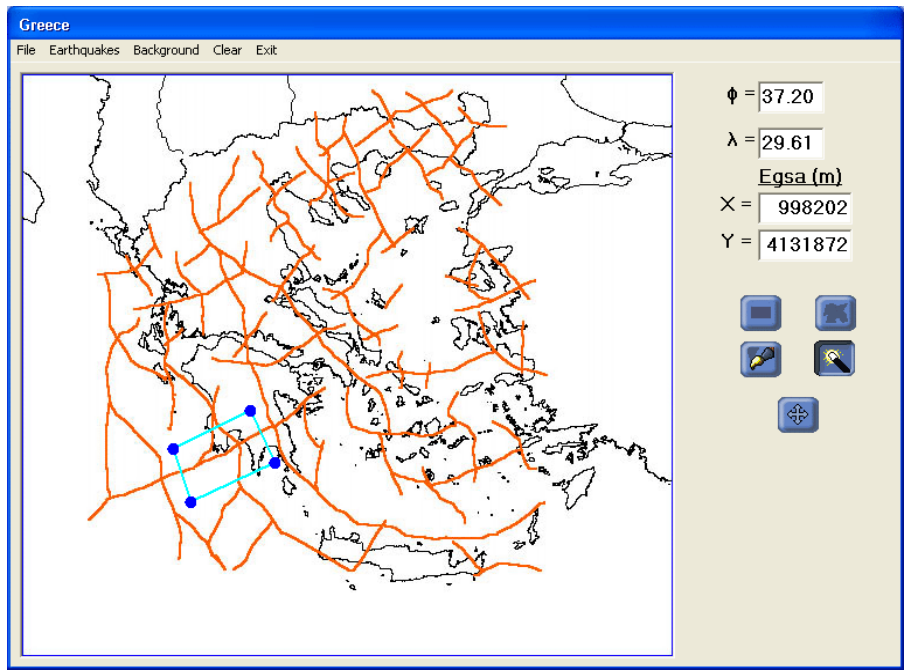

Fig. 14. Regional seismogenic zone (blue frame) which is considered for the application of the lithospheric seismic energy flow model.

For the construction of the cumulative seismic energy release of the specific seismogenic region all the EQs no matter what their magnitude is are taken into account. The sampling interval used is 1 month while the seismic catalog spans back to 1901. In the following figure (15) the calculated cumulative seismic energy release is presented.

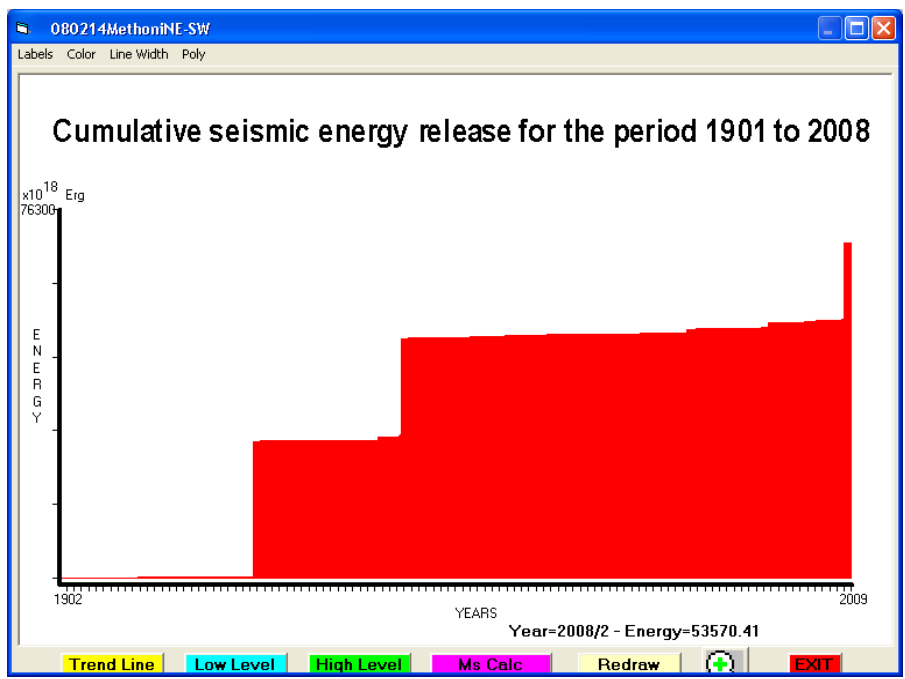

Fig. 15. Cumulative seismic energy release determined for the period from 1901 to 2008. Sampling interval is of one month.

In this diagram the two abrupt steps which are observed are from left to right: the EQ of July 1927, Ms = 7.1R and the EQ of November 1947, Ms = 7R. From 1947 till 2008 no such magnitude EQ took place. The latter means that the released seismic energy was less than what was stored in this seismogenic region. Therefore, a large amount of energy is available for the generation of some future large EQs. The last step which is observed at the right part of the graph corresponds to the Methoni seismic event of 14/2/2008.

Following the lithospheric seismic energy flow model [1] the following figure (16) has been constructed for the maximum expected magnitude of the probable future EQ.

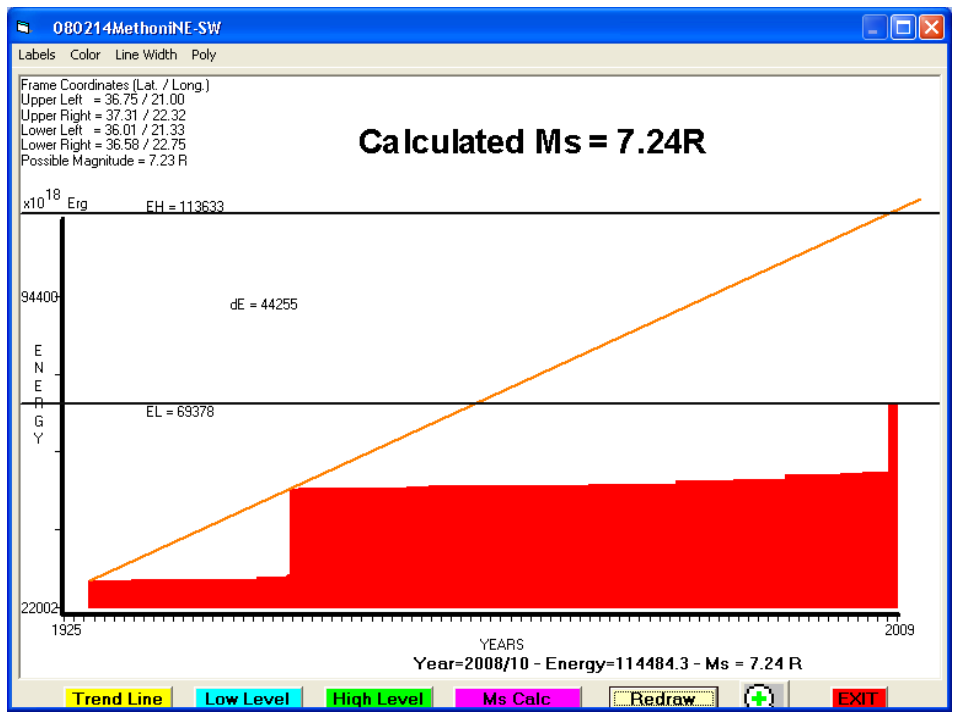

Fig. 16. Determination of the maximum expected magnitude of the probable future EQ in the seismogenic area denoted by the frame of fig. (14). The determined max magnitude is Ms = 7.24R.

The two steps of energy release which are denoted by the EQs of 1927 and 1947, in essence, define the theoretical linear graph of seismic energy release under balanced (incoming energy equals the released one) energy flow through the seismogenic open physical system. This linear graph is represented by the orange straight line which passes from both corners of the steps of the graph. The far right step of the graph indicates the released energy by the Methoni EQ (14/2/2008). Consequently, if it is assumed that an earthquake will take place which will release all the stored seismic energy, in other words the next future step intersects the upper horizontal line, then the magnitude of this EQ will never exceed the value of Ms = 7.24 if it takes place within the first months of 2008. At this point it must be said that the nature is many time unpredictable, so even if smaller magnitude (6 – 7R) EQs are more probable, this seismogenic area has already generated two EQs of magnitude equal or larger than M = 7R.

5. CONCLUSIONS

Summarizing all the results obtained by the application of the various physical models for the Methoni regional seismogenic area what can be said is as follows:

- The Methoni regional seismogenic area is still seismically active.

- Its seismic activity is validated by the presence of the generated electrical field which suggests the seismogenic area as the origin of the generating mechanism of the electric field.

- The Day of occurrence of a probable large EQ has been estimated within a period of 3 days with most probable one the 29$^{th}$ of February of 2008.

- The time of occurrence within the three critical days has been defined by the daily tidal oscillation.

- Its max expected magnitude cannot exceed the value of 7.24R if this EQ takes place within the first few months of 2008. Most probable is a magnitude between 6R and 6.5R leaving a large amount of seismic energy still stored in the seismogenic area.

- Consequently, all the parameters (location, time, magnitude) of a future large EQ have been determined in terms of "short-term prediction" by the use of simple physical models.

- An alternative different and possible option is that the seismogenic region will release slowly seismic energy by any slip mechanism. In this case since the energy release takes a long time period no earthquake will be generated at all.

- What remains more is to see if these results will be verified or falsified by nature.